\documentclass[3p,times]{elsarticle}

\usepackage{ecrc}

\volume{00}

\firstpage{1}

\journalname{Advances in Solar Physics}

\runauth{S. Wedemeyer et al.}

\jid{procs}

\jnltitlelogo{Advances in Solar Physics}

\CopyrightLine{2015}{Published by Elsevier Ltd.}

\usepackage{amssymb}
\biboptions{sort&compress}

\usepackage[figuresright]{rotating}

\def\araa{ARA\&A}%
\def\apj{ApJ}%
\def\apjl{ApJ}%
\def\apjs{ApJS}%
\def\apss{Ap\&SS}%
\def\aap{A\&A}%
\def\aapr{A\&A~Rev.}%
\def\aaps{A\&AS}%
\def\mnras{MNRAS}%
\def\pasj{PASJ}%
\def\solphys{Sol.~Phys.}%
\def\ssr{Space~Sci.~Rev.}%
\def\nat{Nature}%
\def\memsai{Mem.~Soc.~Astron.~Italiana}%
\begin{document}
\vspace*{-30mm}
\ \\ 
\begin{frontmatter}
%% e.g. \dochead{17th International Conference on Dynamical Processes in Excited States of Solids}
\title{\ \\[-3mm]SSALMON - The Solar Simulations for the Atacama Large Millimeter Observatory Network\ \\[-14mm]}

\author[a:ita,a:czarc]{S.~Wedemeyer}
\ead{svenwe@astro.uio.no}

\author[a:nrao]{T.~Bastian}
\author[a:hvar,a:czarc]{R.~Braj{\v s}a}
\author[a:czarc,a:aicz]{M.~Barta}
\author[a:ucb,a:ugla]{H.~Hudson}
\author[a:njit]{G.~Fleishman}
\author[a:stpt,a:mps]{M.~Loukitcheva}
\author[a:esa]{B.~Fleck}
\author[a:ugla]{E.~Kontar}
\author[a:lmsal]{B.~De Pontieu}
\author[a:nasa]{S.~Tiwari}
\author[a:ita]{Y.~Kato} 
\author[a:bale]{R.~Soler}
\author[a:eso]{P.~Yagoubov}
\author[a:chal]{J.~H.~Black}
\author[a:naoj]{P.~Antolin}
\author[a:aicz]{S.~Gunar} 
\author[a:ugla]{N.~Labrosse}
\author[a:fnhw]{A.~O.~Benz}
\author[a:ionn]{A.~Nindos}
\author[a:aip]{M.~Steffen}
\author[a:trin]{E.~Scullion}
\author[a:arma]{J.~G.~Doyle}
\author[a:sgraz]{T.~Zaqarashvili}
\author[a:ugraz]{A.~Hanslmeier}
\author[a:uwar]{V.~M.~Nakariakov}
\author[a:aicz]{P.~Heinzel}
\author[a:colo]{T.~Ayres}
\author[a:aicz,a:czarc]{M.~Karlicky}
\author[]{and the SSALMON Group}

\address[a:ita]{Institute of Theoretical Astrophysics, University of Oslo, Norway}
\address[a:czarc]{European ARC, Czech node, Astronomical Institute ASCR, Ondrejov, Czech Republic}
\address[a:nrao]{National Radio Astronomy Observatory (NRAO), USA}
\address[a:hvar]{Hvar Observatory, Faculty of Geodesy, University of Zagreb, Croatia} 
\address[a:aicz]{Astronomical Institute, Academy of Sciences, Czech Republic}
\address[a:ucb]{UC Berkeley, USA}
\address[a:ugla]{University of Glasgow, UK}
\address[a:njit]{New Jersey Institute of Technology, USA}
\address[a:stpt]{Saint-Petersburg State University, Russia}
\address[a:mps]{Max-Planck-Institut f\"ur Sonnensystemforschung, G\"ottingen, Germany}
\address[a:esa]{ESA Science Operations Department, ESA}
\address[a:lmsal]{Lockheed Martin Solar \& Astrophysics Laboratory, USA}
\address[a:nasa]{NASA's MSFC, USA}
\address[a:bale]{Univ. Illes Balears, Spain}
\address[a:eso]{European Southern Observatory}
\address[a:chal]{Chalmers University of Technology, Dept. of Earth and Space Sciences, Sweden}
\address[a:naoj]{National Astronomical Observatory of Japan (NAOJ), Japan}
\address[a:fnhw]{FHNW, Institute for 4D Technologies, Windisch, Switzerland} 
\address[a:ionn]{Physics Department, University of Ioannina, Greece}
\address[a:aip]{Leibniz-Institut f\"ur Astrophysik Potsdam (AIP), Germany}
\address[a:trin]{Trinity College Dublin, Ireland}
\address[a:arma]{Armagh Observatory, N. Ireland}
\address[a:sgraz]{Space Research Institute of Austrian Academy of Sciences, Austria}
\address[a:ugraz]{Univ. Graz, Inst. of Physics, Austria}
\address[a:uwar]{University of Warwick, UK}
\address[a:colo]{University of Colorado, USA}


\begin{abstract}
\ \\[-6mm]
The Solar Simulations for the Atacama Large Millimeter Observatory Network (SSALMON) was initiated in 2014 in connection with two ALMA development studies. 
The Atacama Large Millimeter/submillimeter Array (ALMA) is a powerful new tool, which can also observe the Sun at high spatial, temporal, and spectral resolution. 
The international SSALMONetwork aims at co-ordinating the further development of solar observing modes for ALMA and at promoting scientific opportunities for solar physics with particular focus on numerical simulations, which can provide important constraints for the observing modes and can aid the interpretation of future observations. 
The radiation detected by ALMA originates mostly in the solar chromosphere -- a complex and dynamic layer between the photosphere and corona, which plays an important role in the transport of energy and matter and the heating of the outer layers of the solar atmosphere.
Potential targets include active regions, prominences, quiet Sun regions, flares. 
Here, we give a brief overview over the network and potential science cases for future solar observations with ALMA. \ \\[-7mm]
\end{abstract}

\textit{Email address: svenwe@astro.uio.no (S. Wedemeyer - corresponding author)}\\[-10mm]


\end{frontmatter}

%\clearpage

% \linenumbers

\section{Introduction}
\label{sec:intro}
%===============================================================================

Observations of the Sun with the Atacama Large Millimeter/submillimeter Array (ALMA) have a large potential for revolutionizing our understanding of our host star. 
The radiation emitted at ALMA wavelengths originates mostly from the chromosphere -- a complex and dynamic layer between the photosphere and the corona, which plays an important role in the transport of energy and matter and the heating of the outer layers of the solar atmosphere.
Despite decades of intensive research, the chromosphere is still elusive and challenging to observe owing to the complicated formation mechanisms behind currently available diagnostics like, e.g., H$\alpha$ and the spectral lines of Ca\,II and Mg\,II. 
The Atacama Large Millimeter/submillimeter Array (ALMA) will change the scene substantially as it serves as a nearly linear thermometer at high spatial, temporal, and spectral resolution, enabling us to study a large range of important scientific topics in modern solar physics \citep{bastian02,2011SoPh..268..165K}. 
In particular, breakthroughs are expected regarding central questions like coronal heating, solar flares and space weather.

ALMA\footnote{Please refer to the ALMA websites for more technical information: http://www.almaobservatory.org/.}  is located on the Chajnantor plateau in the Chilean Andes at an altitude of 5000\,m. 
It consists of 66~antennas, which are sub-divided in two distinct arrays: the \textit{ALMA 12-m Array} consisting of 50~antennas with diameters of 12\,m and the \textit{Atacama Compact Array} (ACA or \textit{``Morita Array''}) with 12  7-m  antennas and 4   12-m Total Power (TP) antennas.  
The TP antennas measure, as the name says, the total power received with a single antenna, which is important for calibration and reconstruction of the point spread function (PSF) of the whole array. 
The combination of the 12-m array, the 7-m antennas (in a fixed compact configuration) and the TP antennas ensures that a large range of spatial scales is sampled, allowing a good characterisation of the PSF. 
However, the 12-m array and the ACA can be used jointly or separately and the 12-m array can even be split into sub-arrays. 
The 12-m array is reconfigurable, with maximum baselines, i.e., the distance between two antennas, as long as 16\,km, yielding an angular resolution less than 10\,mas in the highest frequency bands. 
In practice, solar observing will likely exploit more compact array configurations to more optimally sample the complex and dynamic brightness distribution because the emission fills the whole antenna beam.  
The effective resolution of an extended source like the Sun is then determined by the coverage of the spatial Fourier space (the $u$-$v$-space). 
For ALMA's shortest wavelengths, the spatial resolution of the reconstructed images might be close to the one achieved by current optical solar telescopes, i.e., a few 0.''1 (at a wavelength of $\lambda = 1$\,mm) or better. 
The instantaneous field of view (FOV) is given by the FWHM of the main lobe of a single 12-m antenna and scales with wavelength $\lambda$ approximately as $\theta \approx 19'' \times \lambda/{1\,\mathrm{mm}}$. 
ALMA can therefore only observe  a small fraction of the Sun at a given time but the total FOV can be increased by rapid mosaicing (i.e., multiple pointings) or ``on-the-fly'' continuous scanning.

Every antenna will be equipped with receivers covering up to ten frequency bands in the range from 35\,GHz to 950 GHz (8.6\,mm to 0.3\,mm) although it is not clear yet if/when the receivers for the longest wavelengths will be built. 
ALMA can record data with a high cadence of 1\,s or less and can change between different receivers and thus wavelength bands, which can deliver thousands of spectral channels, both polarised and unpolarised.  
ALMA can safely observe the Sun because the antenna surfaces feature a micro-roughness that scatters most of the intense optical/IR radiation. 
Nevertheless, the large flux density of the Sun poses technical challenges. 
Two strategies for coping with the large solar signal are being considered: 1)~a solar filter can be rotated into the optical path to attenuate the signal; or 2)~the system gain can be reduced by electronically "de-tuning" the front end. 
Test observations of the Sun using both approaches have already been successfully carried out, including the recent campaign \#5 in December 2014. 
Regular solar observations are expected to start in observing Cycle~4 in late 2016.

The opacity at millimeter wavelengths in quiet Sun regions is mainly due to free-free transitions, i.e., thermal bremsstrahlung and H$^-$ \citep[cf., e.g., ][and references therein]{2011SoPh..273..309S,2006A&A...456..697W}. 
The Planckian source function, which is generated under local thermodynamic equilibrium (LTE) conditions, depends essentially linearly on the (local) gas temperature at these wavelengths. 
The continuum intensity or brightness temperature, which can be observed with ALMA, is therefore a rather direct measure of the gas temperatures along the line of sight through the solar chromosphere. 
This property turns ALMA into a linear thermometer for the plasma in the solar chromosphere, which is in principle much easier to interpret than other chromospheric diagnostics in the ultraviolet, visible or infrared range. 
Another interesting property results from the fact that the effective formation height of the  continuum radiation at millimeter wavelengths increases with height \cite{2007AnA...471..977W,2015A&A...575A..15L}.  
At $\lambda \approx 0.3$\,mm, which is the shortest wavelengths accessible by ALMA, the continuum intensity originates from the layer between the high photosphere and the low chromosphere (i.e., around the classical temperature minimum region). 
At the longest available wavelength, which will possibly be $\lambda \approx 3.6$\,mm, ALMA essentially maps the upper chromosphere. 
Observing the solar chromosphere at different wavelengths would therefore, in principle, allow for mapping different layers in rather rapid sequence. 
Utilizing this effect for yet to be developed tomographic techniques may enable ALMA to probe the time-dependent three-dimensional thermal structure of the solar chromosphere \cite{2007AnA...471..977W}. 
In addition, there is a non-thermal contribution to the intensity at millimeter wavelengths, which is due to synchrotron radiation generated by high-energy electrons. 
This non-thermal component will be significant during flares and can be observed with ALMA.

ALMA provides further diagnostic possibilities, which will result in important findings for the solar chromosphere. 
The polarisation of the measured radiation allows the determination of the magnetic field in the continuum forming layers, i.e. at different heights of the solar chromosphere \citep{2015A&A...575A..15L}. 
As mentioned above, each receiver band can contain up to a few thousand spectral channels so that ALMA can provide rapid sequences of spectral data cubes. 
It will enable the study of radio recombination lines (``Rydberg transitions'') and possibly molecular lines of carbon monoxide \citep[see, e.g.,][and references therein]{1981ApJ...245.1124A,2005A&A...438.1043W,2006ApJS..165..618A} and other simple molecules. 
Such spectral lines may bear further information on plasma properties such as gas temperature, gas pressure and velocity. 
However, such lines have not been observed on the Sun at this high spatial resolution yet so that appropriate diagnostic tools have to be developed in order to exploit the full potential of this promising option.

%================================================================================
\begin{figure}[pt!]
\centering
\resizebox{15cm}{!}
{\includegraphics[]{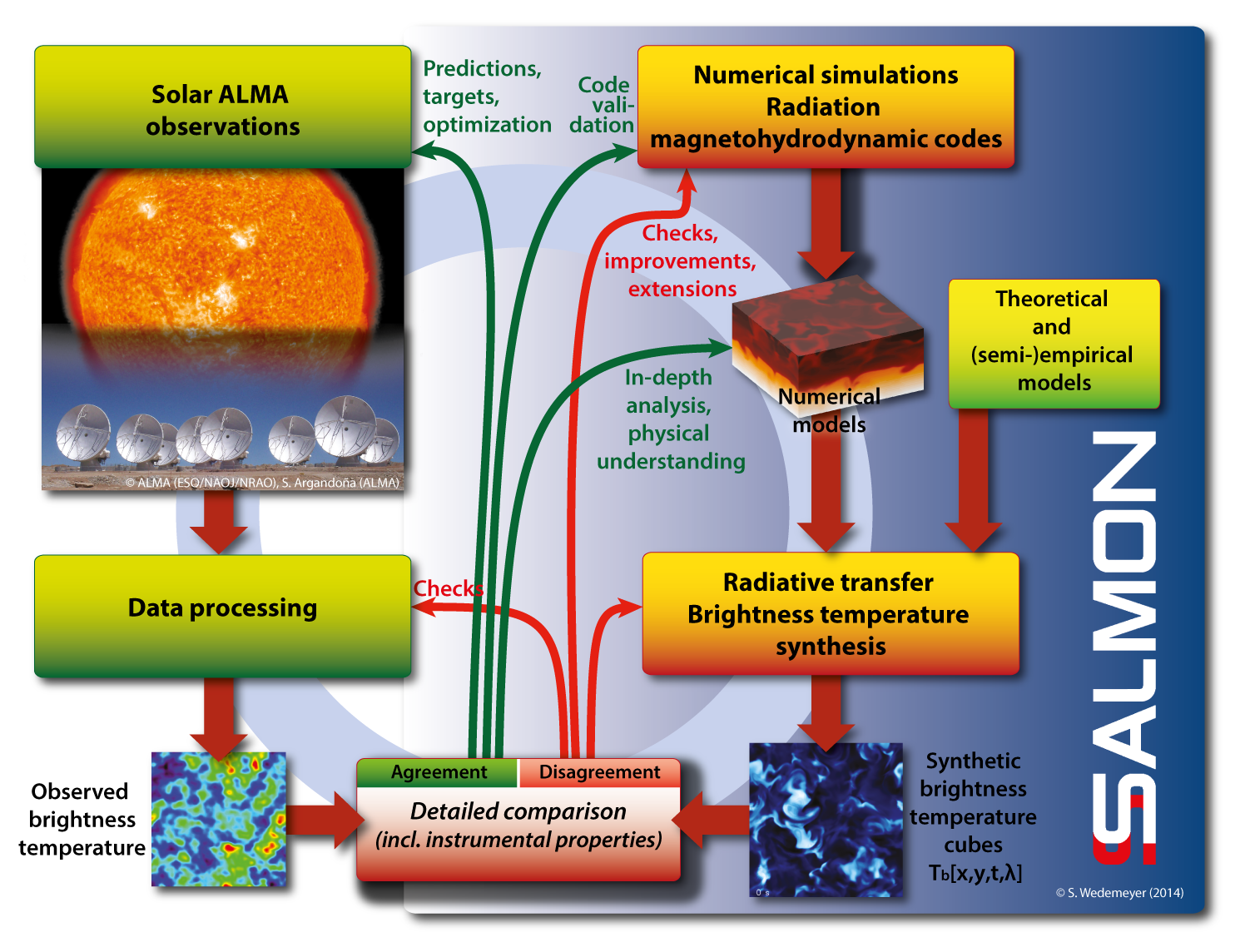}}
\caption{Detailed comparisons of ALMA observations (left) with numerical models (right) enable us to develop observing strategies for ALMA, to plan, optimize, and interpret observations, and to demonstrate that potentially important scientific results can be expected from solar ALMA observations. 
}
\label{fig:scheme}
\end{figure}
%================================================================================

ALMA's unprecedented capabilities at millimeter wavelengths promise significant new findings for a large range of topics in solar physics including the dynamics, thermal structure and energy transport in the ÒquietÓ solar chromosphere, active regions and sunspots, spicules, prominences and filaments, and flares.
Entering a new domain in terms of spatial resolution for millimeter wavelengths also requires that new observing strategies and diagnostic tools are to be developed. 
State-of-the-art numerical models of the solar atmosphere are very useful in this respect.  
They can play an important role for the planning, optimizing and interpretation of observations with ALMA. 
Synthetic brightness temperature maps, which are calculated from numerical models or which are based on other chromospheric diagnostics, can be used to simulate what ALMA would observe. 
Different instrumental set-ups can be tested and adjusted to the scientific requirements, finding the optimal set-up for different scientific applications. 
The general procedure includes tasks like calculating models of the solar atmosphere, producing synthetic brightness temperature maps, applying instrumental effects and comparisons with real ALMA observations of the Sun (see Fig.~\ref{fig:scheme}). 
The individual steps are demanding in themselves but are in principle similar for a large range of scientific targets. 
In order to use these synergies, which will enable the solar physics community to use ALMA in an optimum way, we initiated a scientific network named Solar Simulations for the Atacama Large Millimeter Observatory Network (SSALMON). 
In this article, we outline the scientific infrastructure that led to the formation of SSALMON (Sect.~\ref{sec:scinfra}) and give a brief overview over some potential science cases for ALMA observations of the Sun (Sect.~\ref{sec:sciencecases}). 
A more extensive review article is currently in preparation \cite{ssalmon_ssrv15}.

%================================================================================
%================================================================================
%================================================================================
\section{Scientific infrastructure} 
\label{sec:scinfra}
%================================================================================

Solar observing is an important component of the ALMA science program and the instrument has been designed to support such observations (see Sect.~\ref{sec:intro}). 
However, solar observing at millimeter wavelengths has unique challenges that require development well beyond hardware needs. 
ALMA is currently the largest international astronomy facility of the world, which was achieved by a partnership of Europe, North America and East Asia in cooperation with the Republic of Chile.
It is operated by a Joint ALMA Observatory (JAO) and ALMA Regional Centers (ARCs, see Sect.~\ref{sec:arcnodes} and Fig.~\ref{fig:ARCs}). 
In order to enable ALMA's use by the broader international solar community, two development studies have been funded -- the North American ALMA solar development program (see Sect.~\ref{sec:almastudy_na}) and the ESO (European Southern Observatory) ALMA solar development program (see Sect.~\ref{sec:almastudy_eu}). 
A major goal is to coordinate effort by the members of the international solar physics community to develop recommendations and requirements for implementing an initial suite of solar observing modes for use by the wider community on ALMA in \mbox{Cycle~4}.
On September 1st, 2014, the Solar Simulations for the Atacama Large Millimeter Observatory Network (SSALMON) was initiated in connection with these two international ALMA development studies and in collaboration with the Czech node of the ALMA Regional Center (see Sect.~\ref{sec:alma_czecharc}). 
The main purpose of the network is to co-ordinate activities related to solar science with ALMA and to promote the scientific potential of ALMA observations of the Sun with particular focus on modeling aspects.
More details are given in Sect.~\ref{sec:ssalmon}.

%================================================================================
\subsection{ALMA Regional Centers and the role of the Czech node for solar physics} 
\label{sec:alma_czecharc}
\label{sec:arcnodes}

\begin{figure}[t]
\centering
\resizebox{12cm}{!}
{\includegraphics[]{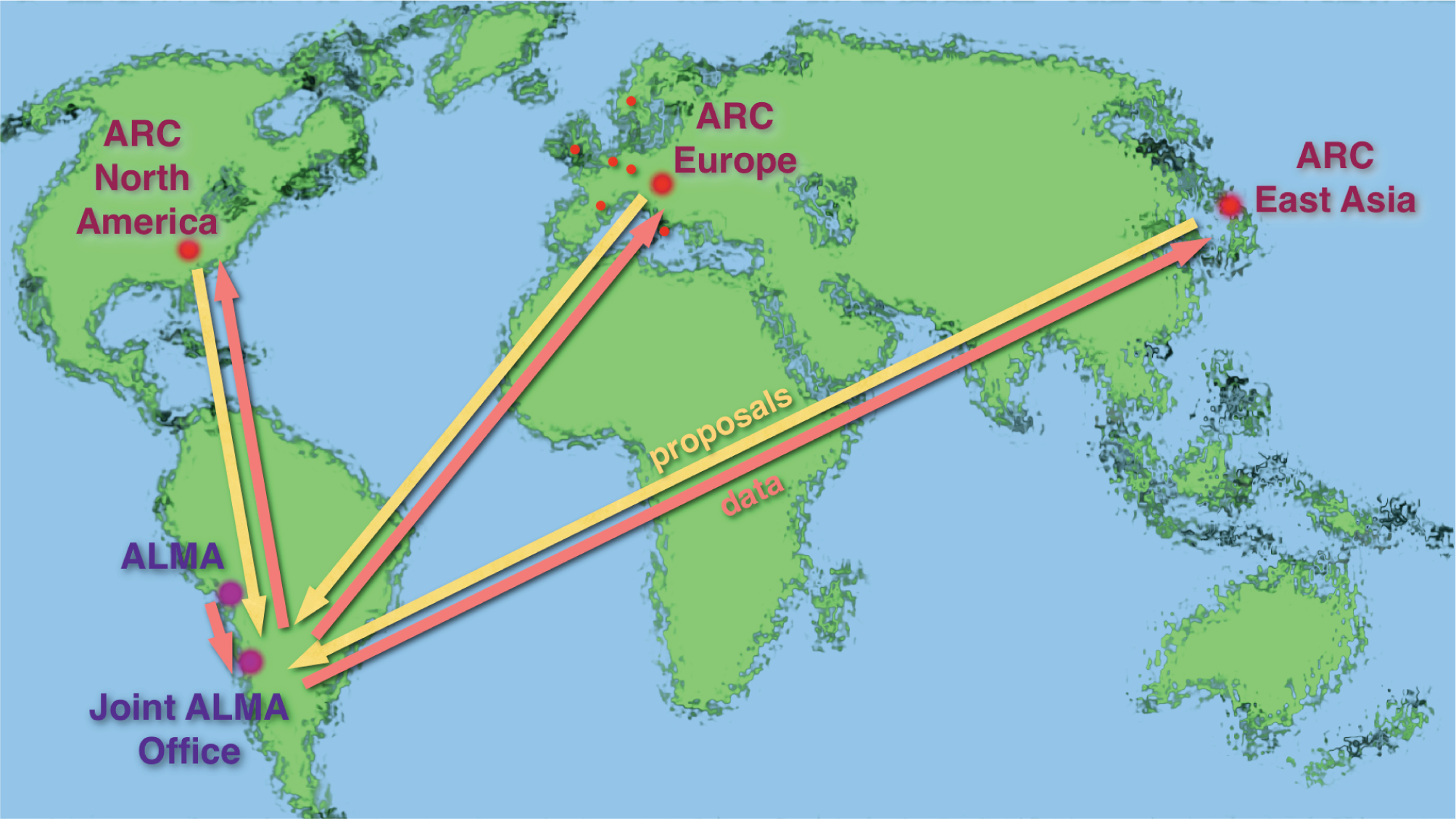}}
\caption{European, North American and East Asian ALMA Regional Centres -- an
  infrastructure for ALMA user community.} 
\label{fig:ARCs}
\end{figure}

ALMA is a unique cutting-edge research facility and a lot of money and effort has been invested into its construction and development. 
It is therefore important to ensure that it will be used for projects of highest scientific excellence. 
At the same moment -- because of the complexity of ALMA -- just a small fraction of the potential research community has acquired sufficient technical expertise to be able to propose projects that would fully take advantage of ALMA capabilities. 
Hence, even the greatest scientific ideas could be lost because of unawareness of their proponents about ALMA or lack of required technical knowledge. 
In order to increase accessibility of ALMA to a much broader scientific community the Joint ALMA Observatory created a support infrastructure -- the network of ALMA Regional Centers (ARCs):  
The European (EU)~ARC operated by ESO, the North American (NA)~ARC run by the National Radio Astronomy Observatory (NRAO), and the East Asia (EA)~ARC managed by the National Astronomical Observatory of Japan (NAOJ) - the partner institutions that participate in the ALMA construction, development and operations (Fig .~\ref{fig:ARCs}).

The main role of ARCs is to serve as an interface layer between the ALMA observatory and the research community. 
In order to accomplish this task, the ARCs and their nodes (see below) provide the following:

\begin{itemize}

\item 
  Provide support (also personal - face-to-face/F2F) to the members of the research 
  community in proper usage of ALMA in all stages of their research project
  (help with proposal submission, negotiations with the ALMA astronomers on
  technical details of the project, scientific data reduction and quality
  assurance, etc.).

\item 
  Spread the technical knowledge and awareness of ALMA among the research
  community (workshops/schools, training).

\item
  Help the ALMA developers (e.g., tests of infrastructure and user software, laboratory
  molecular spectroscopy - updates of spectral line catalogues).

\item
  Promote and define user-community driven enhancements of ALMA (e.g., new
  observing modes).

\end{itemize}

%================================================================================
%================================================================================

Unlike the NA and EA~ARCs, the European ARC is not a compact infrastructure but it has been formed as a distributed network of nodes centered around the ESO headquarters in Garching, Germany. 
One of the seven EU ARC nodes has been established in 2011 at the Astronomical Institute of the Academy of Sciences of the Czech Republic (AI~ASCR) in Ondrejov, Czech Republic. 
Main advantages of the coordinated distributed network of the ARC nodes can be seen in the complementarity of its research expertise, direct involvement of multiple cooperating European research institutions from several countries, and geographical (and possibly lingual) proximity of the RIsâ nodes  to their clients (i.e., the ALMA users). 
The coordinated-network model adopted for the EU ARC and its main strong points are described in \cite{2014SPIE.9149E..0YA}.

Clearly, one of the prominent features of the European distributed ARC is the diversity in research expertise among the nodes. Interested readers are referred to the European ARC Guide document\footnote{Available on-line at
{http://almascience.eso.org/documents-and-tools/cycle-2/alma-eu-arcguide}.} for more details. 
The Czech node of the EU~ARC has a unique expertise, namely solar radio physics, and has been delegated to be the main support facility for the entire European user community in solar research with ALMA until the solar observing mode will be commissioned. 
Let us note that, in addition to this future services, it undertakes already now also support of the projects in other fields of its expertise (galactic and extragalactic astrophysics, laboratory molecular line spectroscopy) for the broader region of Central and Eastern Europe, because of its geographical location. 
The NA and EA community are served by their respective ARCs, whereas the support of solar physicists in Chile will beâ similar as in other fields of ALMA observationsâ shared among the EU (Ondrejov node), NA and EA~ARCs.

Although the Sun is one of the envisaged ALMA scientific targets, regular solar observations have not yet started because of the many technical challenges that must be addressed  (the large radio flux density, short dynamical scales, calibration issues, source tracking, etc). 
These issues have to be resolved before the solar ALMA observing mode can be commissioned. 
Two studies, one funded by the NSF in North America and one funded by ESO in Europe, are coordinating efforts to address these issues.
Based on its specific expertise in this area, the Czech node was delegated by ESO to carry out the project \textit{``Solar Research with ALMA''} in cooperation with the NA and EA ARCs (see
Sect.~\ref{sec:almastudy_eu} for details).

%================================================================================
\subsection{The North-American led ALMA study}
\label{sec:almastudy_na}
%================================================================================

\textit{``Advanced Solar Observing Techniques''} is a project within the North American Study Plan for Development Upgrades of the ALMA. 
The Principal Investigator (PI) of this project is T.~Bastian, NRAO, USA.
North American study (funded by the National Science Foundation, NSF) is comprised of thirteen collaborators representing all of the ALMA international partners, as well as more than thirty affiliated scientists. 
The North American team was launched in April 2014.
The constituent activities of the North American study are:

\begin{itemize} 
\item Definition of single dish and interferometric observing modes, supported by simulations.
\item Development of detailed use cases, supported by theoretical modeling as appropriate.

\item Commissioning and science verification activities designed to verify instrument performance, develop data calibration techniques, and test specific observing strategies.
\item Development of requirements for online and offline software support of solar observing modes and solar data calibration and reduction.
\item Development of recommendations to the ALMA project for solar observing modes to be supported during Cycle~4.
\end{itemize} 

The science simulations group within the study formed the basis for  
SSALMON, which was rolled out in September~2014 (see Sect.~\ref{sec:ssalmon}). 
The North American team participated in a major commissioning and science verification campaign in December 2014. 
Another team activity concerns an ALMA solar science workshop, which is planned for sometime late in 2015.

%================================================================================
\subsection{The European led ALMA study} 
\label{sec:almastudy_eu}
%================================================================================

ALMA is funded in Europe by the European Organization for Astronomical Research in the Southern Hemisphere (ESO). 
In November 2014, a new project, \textit{``Development Plan Study: Solar Research with ALMA''}, in the frame of the ESOÕs Proposal \textit{``Advanced Study for Upgrades of the Atacama Large Millimeter/Submillimeter Array (ALMA)''}, was started. 
The project is carried out at the Czech ALMA ARC node, based in the Astronomical Institute of the Academy of Sciences of the Czech Republic, Ondrejov (see Sect.~\ref{sec:alma_czecharc}) and will last for 30~months. 
The Principal Investigator of the Project is R. Braj{\v s}a (Hvar Observatory, Faculty of Geodesy, University of Zagreb, Croatia), while the Project Manager is M. Barta (Czech ALMA ARC node, Ondrejov, Czech Republic). 
The overall aim of the project is to provide an in-depth understanding of the requirements for solar observing with ALMA. 
The specific elements are:  
\begin{itemize} 
\item Development of a set of detailed use cases for solar observing with ALMA: \\
The use cases include the quiet Sun and dynamic chromosphere, filaments on the disc and prominences at the limb, active regions and sunspots, solar flares and molecular and recombination spectral lines. 
\item Definition of solar observing modes for ALMA: \\
The observing modes embrace single-dish and interferometric observations of the Sun.   
\item Analysis of the calibration requirements for these modes
\item Identifying software requirements for observing preparation, execution and data reduction: \\
ALMA software will be adapted and modified taking into account specific requirements for solar observations. 
\item Modelling of the radiation for various solar structures in the ALMA wavelength range is also an important part of this project, with the aim to compare calculation results with real data obtained by observations. 
\end{itemize}

It is expected that this study will be performed in close collaboration with the equivalent North American Development Study (see Sect.~\ref{sec:almastudy_na}) and with related activities in East Asia. 
An important part of the project are test campaigns for solar observations with ALMA, such as the commisioning and science verification (CSV) campaign which took place in December 2014. 
The final goal of this study is to help implementing selected solar observing modes within ALMA observing Cycle 4, starting with Bands~3 ($\nu = 84-116$\,GHz, $\lambda = 2.6 - 3.6$\,mm) and 6 ($\nu = 211 - 275$\,GHz, $\lambda = 1.1 - 1.4$\,mm).

%================================================================================
\begin{figure}
\centering
\resizebox{11.5cm}{!}
{\includegraphics[]{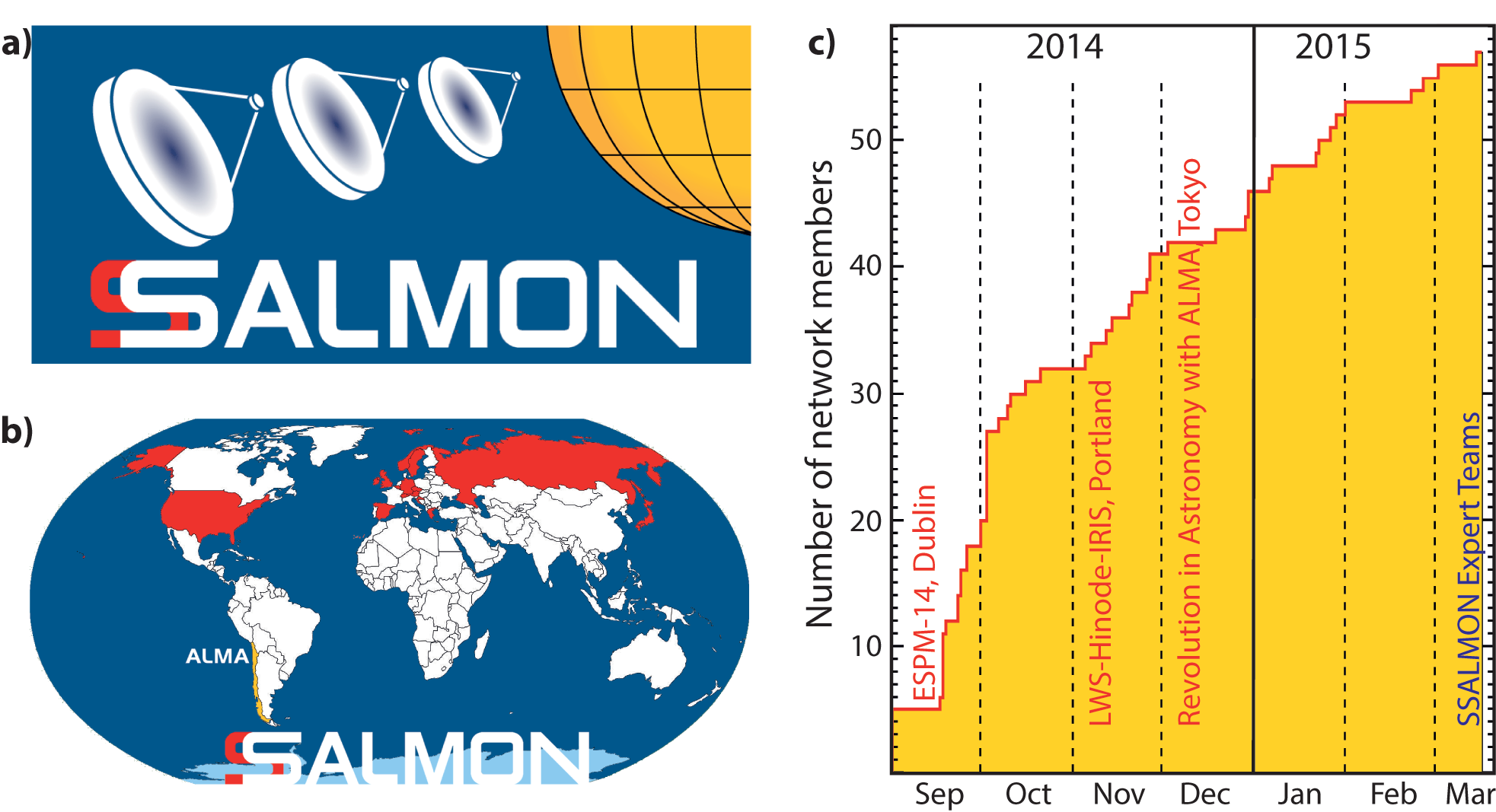}}
\caption{a) The official logo of Solar Simulations for the Atacama Large Millimeter 
Observatory Network, depicting three idealised antennas and the Sun. 
The background color and font are inspired by the ALMA logo.
b)~First affiliations of all network members marked on a world map. 
c)~Number of SSALMON members since the network started on September 1st, 2014 (as of December 31st, 2014). Three major conferences, at which SSALMON was promoted, and the start of the SSALMON Expert Teams are marked.}
\label{fig:ssalmon}
\end{figure}
%================================================================================

%================================================================================
\subsection{The Solar Simulations for the Atacama Large Millimeter Observatory Network}
\label{sec:ssalmon}

As mentioned above, the Solar Simulations for the Atacama Large Millimeter Observatory Network (SSALMON) was initiated on September 1st, 2014, by the network co-ordinator S.~Wedemeyer (University of Oslo, Norway) in connection with the NA and EU development studies and in collaboration with the Czech node of the ALMA Regional Center. 
As of March~17th, 2015, 57~scientists from 15~countries (plus ESA and ESO) have joined the network.  
The SSALMON  logo is shown in Fig.~\ref{fig:ssalmon}a together with a world map (Fig.~\ref{fig:ssalmon}b), in which all countries of the primary affiliations of all members are marked. 
The growth of the network in terms of number of members is plotted in Fig.~\ref{fig:ssalmon}c.

The activities of the network focus on all related simulation and modelling aspects ranging from calculating models of the solar atmosphere, producing synthetic brightness temperature maps, applying instrumental effects, comparisons with real ALMA observations of the Sun to developing optimized observation strategies. 
The aims of the SSALMONetwork can be summarised as follows: 

\begin{itemize} 
\item \textbf{Key goal 1:} Raising awareness of scientific opportunities with ALMA. Simulations can demonstrate what could be possibly observed with ALMA and which scientific problems could therefore be addressed in the future, also in combination with other ground-based and space-borne instruments.
\item \textbf{Key goal 2:} Clear visibility of solar science within the ALMA community. A demonstration of potentially important scientific results will help ensure a fair share of the observing time for solar campaigns.
\item \textbf{Key goal 3:} Constraining ALMA observing modes through modeling efforts. Simulations in comparison with ALMA observations will help to develop optimal observation strategies and to plan and analyze solar observations.
\end{itemize}

This overview article is one of the first major network activities. 
It will be followed by a more extensive review article in 2015 \cite{ssalmon_ssrv15}, which will give an overview over possible science cases and big questions in solar physics to be investigated with ALMA. 
A short summary is provided in Sect.~\ref{sec:sciencecases}. 
Based on these publications, a number of expert teams will be formed in early 2015. 
Each team will work on a specific task in preparation of future regular ALMA observations and their interpretation. 
Common strategies and tools will be developed and provided to all teams. 
A workshop or small conference will be organised possibly still in 2015 in co-operation with two development study teams and the Czech ARC node.

The network is open to everybody who has a professional interest in contributing to potential ALMA solar science, which include or require simulations.
More information and a registration form can be found on the network pages, which are hosted by the University of Oslo, Norway: http://ssalmon.uio.no. 

%================================================================================
%================================================================================
%================================================================================
\section{Potential science with ALMA as predicted by numerical simulations} 
\label{sec:sciencecases}
%================================================================================
%================================================================================
%================================================================================
%================================================================================

ALMA will advance our knowledge about the solar atmosphere in a large range of different aspects ranging from quiet Sun regions to solar flares. 
In particular, substantial progress can be expected regarding (i)~coronal and chromospheric heating, (ii)~solar flares, and (iii)~space weather, which are central topics in modern solar physics. 
In addition, ALMA will be able to contribute significantly to essentially all topics that concern the solar chromosphere, either directly or indirectly.  
A brief, non-extensive overview over potential science cases is given in the following sections. 
A comprehensive overview is currently in preparation \cite{ssalmon_ssrv15}.

%================================================================================
%================================================================================
%================================================================================
\subsection{Coronal and chromospheric heating} 
\label{sec:heating}
%================================================================================

ALMA probes the (3D) thermal (and magnetic) structure and dynamics of the solar chromosphere and thus sources and sinks of energy.  
Observations with ALMA may thus reveal the signatures of atmospheric energy transport and heating mechanisms and thus help to identify the  most relevant processes, which would be substantial progress towards solving the question about how the outer layers of the Sun are heated. 
It is important to emphasise here that the layers of the solar atmosphere are intricately connected \cite{2009SSRv..144..317W} so that ALMA observations of the chromosphere have direct implications for heating of the layers above, i.e. for the coronal heating problem, which is one of the big standing problems of astrophysics today.

So far it is still not entirely clear what mechanisms are the most dominant sources of atmospheric heating although a large number of different processes have been identified, which could potentially provide and dissipate energy in the chromosphere and corona \cite[see, e.g.,][and references therein]{Schroeder2012,Cuntz2007,Judge2003,Judge1998,Buchholz1998,Schrijver1995}. 
Among them are mechanisms connected to spicules \citep{2007PASJ...59S.655D, 2006ASPC..354..276R}, Alfv\'en waves \cite{2007Sci...318.1574D,2009Sci...323.1582J,2012Natur.486..505W,2014Sci...346D.315D,2010ApJ...711..164V,2014ApJ...796L..23S}, magneto-acoustic shocks \citep{2006ApJ...648L.151J, 2006ApJ...647L..73H,2008A&A...479..213B}, gravity waves \citep{2008ApJ...681L.125S}, high frequency waves \cite{2008ApJ...680.1542H}, multi-fluid effects \cite{2010ApJ...724.1542K,2012ApJ...753..161M},  plasma instabilities \cite[e.g.,][]{2008A&A...480..839F}, magnetic reconnection and (nano)flare heating~\cite{2008LRSP....5....1B}.
Furthermore, transverse magnetohydrodynamic (MHD) waves, which have been shown to be ubiquitous in the solar atmosphere \cite{2007Sci...318.1574D,2011Natur.475..477M,2007Sci...318.1577O,2007Sci...317.1192T}, are potentially important heating agents along  chromospheric / coronal waveguides \cite{1978ApJ...226..650I,1991SoPh..133..227S,2008ApJ...679.1611T,2008ApJ...682L.141A,2010ApJ...718L.102V} such as coronal loops, which are rooted in the network and plage regions in the layers below~\cite{1977ARA&A..15..363W}.
Such waves have been predicted to trigger current sheets and vortices via resonant absorption and Kelvin-Helmholtz instabilities, which generates a fine strand-like structure of magnetic loops \cite{2014ApJ...787L..22A,2008ApJ...687L.115T}.
High-resolution observations with ALMA can help to investigate important questions in this respect, regarding the generated fine structure and the dissipation mechanisms of these Alfv{\'e}nic waves and thus the heating contribution of transverse MHD waves. 
Estimations of the heating rate due to Alfv\'en waves damped by ion-neutral collisions (\cite{2011JGRA..116.9104S,2011ApJ...735...45G,2013ApJ...777...53T})  suggest that this mechanism may generate sufficient heat to compensate the radiative losses at low altitudes in the solar atmosphere \cite[see also][]{2015A&A...573A..79S}.
On the other hand, \cite{2013A&A...549A.113Z} investigated the role of neutral helium and stratification on the propagation of torsional Alfv\'en waves through the chromosphere. While high-frequency waves are efficiently damped by ion-neutral collisions, \cite{2013A&A...549A.113Z}  concluded that low-frequency waves may not reach the transition region because they may become evanescent at lower heights owing to gravitational stratification. 
Hence, propagation of  Alfv\'en waves through the chromosphere into the solar corona should be considered with caution. 
The ability of ALMA to observe different layers in the chromosphere by using different bands, along with the high temporal and spatial resolutions,  may be crucial to understand how magnetohydrodynamic waves actually propagate and damp in the chromosphere and thus how they contribute to chromospheric heating.

Magnetohydrodynamic waves are also ubiquitously observed in solar prominences \citep{2009ApJ...704..870L,2013ApJ...779L..16H}. 
Since the properties of the prominence plasma are akin to those in the chromosphere, dissipation of wave energy due to ion-neutral collisions may also play a role in the energy balance of prominences. 
In the same way as in the chromosphere, ALMA observations may shed light on the impact of high-frequency waves on prominence plasmas (see also Sect.~\ref{sec:promi}).

%================================================================================
%================================================================================
%================================================================================
\subsection{Solar flares} 
\label{sec:flares}
%================================================================================

Flares produce transient enhancements of emitted radiation over a very extended wavelength range on top of a more slowly varying background emission \cite[e.g.][]{1965sra..book.....K}. 
The emission at radio and micro-wave wavelengths is produced by a number of mechanisms including  gyrosynchrotron emission by 
flare-accelerated electrons and thermal free-free emission. 
Still there are many open questions concerning flares. 
For instance, the physical mechanism behind an observed emission component at sub-THz frequencies remains unclear 
\cite{2001ApJ...548L..95K,2002A&A...381..694T,2004A&A...420..361L,2004ApJ...603L.121K,2008A&A...492..215C,2010ApJ...709L.127F,2013A&ARv..21...58K,2014SoPh..289.3017Z,2014ApJ...791...31K}. 
ALMA's  spectral and spatial resolution for observations in the sub-THz range promise ground-breaking discoveries in this respect.

In general, ALMA will provide invaluable spectral information on flares, especially in the lower solar atmosphere where intense energy release occurs in ribbon and footpoint sources \cite[e.g.,][]{2011SSRv..159...19F,2012ApJ...753L..26M}.
We have never had any access to the continuum formed in these regions, and expect that the ALMA observations will immediately provide powerful new insights. 
Recent observations at 30\,THz provide hints about the novelty of such observations \cite{2013ApJ...768..134K}. %[Kaufmann et al. 2014].
The non-thermal spectra of solar flares are the key to recovering the fast particle spectrum and other physical parameters of flaring loops \cite{2013SoPh..288..549G}. 
This information is vitally needed to investigate the efficiency of the particle acceleration in flares and, thus, to determine the responsible acceleration mechanism(s). 
At those high frequencies the gyrosynchrotron emission is produced by relativistic particles, which can be either accelerated electrons or secondary positrons (created in nuclear interactions). 
In order to distinguish the positron and electron contributions, one has to measure the sense of circular polarization together with the direction of the magnetic field vector (upward or downward) to determine the magneto-ionic mode dominating the emission \cite{2013PASJ...65S...7F}. 
So far, only indirect identifications of the wave mode are available with the Nobeyama data
\cite{2013PASJ...65S...7F}. 
ALMA with its broad spectral range, high spatial resolution, polarization measurement purity (combined with the superior vector magnetic measurements of the Helioseismic and Magnetic Imager (HMI) \cite{2012SoPh..275..207S} onboard the Solar Dynamics Observatory (SDO) \cite{2012SoPh..275...17L}) will be an ideal tool for detecting the relativistic positron (and, thus, nuclear) component in solar flares, providing a powerful but almost unexplored diagnostics.

Another intensively investigated physical phenomenon are quasi-periodic pulsations (QPP) in flares~\cite{2009SSRv..149..119N}. 
Discrimination between several proposed mechanisms (e.g., MHD oscillations at flaring sites and nearby; oscillatory regimes of magnetic reconnection), and hence creating new tools for the diagnostics of flaring plasmas requires the combination of spatial, time and spectral resolution, to be provided by ALMA.

%================================================================================
%================================================================================
%================================================================================
\subsection{Space weather} 
%================================================================================

An important aspect of solar prominences (see Sect.~\ref{sec:promi}) is that they can exist over a wide variety of latitudes from inside active regions (see Sect.~\ref{sec:ar}) all the way to the polar crown, so they are a truly global solar phenomenon. 
For reasons that are still not fully understood, they can suddenly loose equilibrium and erupt, sending huge amounts of magnetised plasma in the interplanetary space. 
Solar prominences are closely related to two of the most violent phenomena found in the solar system, namely, solar flares (see Sect.~\ref{sec:flares}) and Coronal Mass Ejections (CMEs). 
Hence, prominence eruptions are often associated to substantial perturbations of the space environment in the heliosphere, including around the Earth, planets and other solar-system bodies. 
This is what we call ``Space Weather''. 
Due to these effects, understanding the formation of solar filaments and subsequently the origin of solar flares and CMEs is of the utmost importance.
With its high spatial and temporal resolutions, coupled to a large range of available (and quite unexplored) wavelengths,  ALMA is sure to provide an invaluable insight as to the role and impact of filaments and prominences on the evolution of the physical conditions in our environment in Space.

%================================================================================
%%================================================================================
%\subsubsection{Radio Recombination Lines}
%\task{[J. H. Black ( + T. Ayres?): submm-wave Rydberg transitions (recombination lines),
%comments about a few other types of atomic transitions; Are molecular lines (CO from the COmosphere) are detectable at submm wavelengths?] }
%
%%================================================================================
%================================================================================
\subsection{Small-scale structure and dynamics of quiet Sun regions and chromospheric waves}

Quiet Sun regions, i.e., regions outside strong magnetic field concentrations, are in principle easier to model than active regions (see Sect.~\ref{sec:ar}) with the numerical tools available today although quiet Sun regions are complex and not yet understood in all detail.  
It nevertheless makes them an important test case and a potential means of calibration for ALMA observations of the solar chromosphere. 
Quiet Sun models have developed from classical static model atmospheres like, e.g., the one-dimensional semi-empirical model atmospheres by Vernazza, Avrett \& Loeser (VAL) \cite{1981ApJS...45..635V} over detailed one-dimensional dynamic simulations (e.g., the time-dependent simulations by Carlsson \& Stein (CS) \cite{1994chdy.conf...47C,1995ApJ...440L..29C}) towards 
three-dimensional radiation magnetohydrodynamics simulations \cite{2000ApJ...541..468S,2004A&A...414.1121W,2005ESASP.592E..87H,2005ESASP.596E..65S,2010MmSAI..81..582C,2012Natur.486..505W}. 
These different types of models have been used to predict how quiet Sun regions would appear at millimeter wavelengths \cite{2004A&A...419..747L,2007AnA...471..977W,2015A&A...575A..15L}.

The calculations of the emergent thermal free-free radiation at sub-mm and mm wavelengths for the CS simulations produced brightness temperatures that vary strongly in time due to the propagating shock waves and oscillation modes
%but provide a much more reliable proxy for the local gas temperature in the radiation forming layers 
 \cite{2004A&A...419..747L}. 
%
%Wave periods of approximately 3 min can be clearly distinguished in the intensity at all considered wave- lengths. 
%as expected for chromosphere 
The resulting synthetic radiation was compared to interferometric observations of quiet Sun regions with the Berkeley-Illinois-Maryland Array (BIMA) at a wavelength of 3.5\,mm \cite{2006A&A...456..713L,2008Ap&SS.313..197L} although the small-scale pattern was not resolved in these observations. 
The dynamic picture was confirmed by the studies based on 3-D models \cite[e.g.,][]{2007AnA...471..977W}, which in addition exhibited an intermittent and fast evolving pattern in the synthetic millimeter continuum intensity maps 
with structures on scales down to 0.1\,arcsec and time scales on the order of a few 10\,s. 
This pattern correlates closely with the thermal pattern at chromospheric heights as it was already known from earlier simulations \cite{2000ApJ...541..468S,wedemeyer03c,wedemeyer03a}. 
The pattern is produced by propagating shock waves like they were seen in the 1-D simulations by Carlsson \& Stein \cite{1994chdy.conf...47C}. 
The co-existence of hot gas in shock fronts and cool gas in post-shock regions provides an explanation for the observation of carbon monoxide, which is incompatible with the high temperatures in VAL-type models \cite{ayres96,asensio03,2005A&A...438.1043W,2007A&A...462L..31W}. 
Observations of continuum radiation, radio recombination lines and molecular lines of CO (see Sect.~\ref{sec:intro}) will provide crucial tests for the existing models and their further development.

\paragraph{Wave propagation in the solar atmosphere} 
Waves and oscillations are interesting not only from the point of view that they can transport  energy into the chromosphere
(see Sect.~\ref{sec:heating}), they also serve as probes of the structure of the atmosphere in which they propagate \cite[see, e.g.,][and references therein]{Fleck1989}. 
%
%ALMA offers the unprecedented opportunity to study high-frequency waves in the chromosphere.
The high temporal resolution of ALMA will enable measurements of chromospheric oscillations and propagating waves \cite[see, e.g.,][]{2004A&A...419..747L,2006A&A...456..713L,2008Ap&SS.313..197L,2006A&A...456..697W,2007AnA...471..977W} from which information about the chromospheric medium can be extracted. 
Multi-wavelength time series of ALMA observations facilitate travel time measurements between different heights as these disturbances propagate through the chromosphere and thus should finally settle the long-standing question about the propagation characteristics of acoustic waves in the chromosphere.

%================================================================================
\subsection{Strong magnetic fields -- Active regions, sunspots, and coronal loops} 
%\task{[G.~Fleishman: Active region modelling and predictions for ALMA]}
%================================================================================

\paragraph{Active regions and sunspots}
\label{sec:ar}
Modern ground and space-based telescopes have revealed the presence of a variety of small-scale dynamic events in and above sunspot penumbrae, namely, running penumbral waves, dark lanes on penumbral filaments, inward motion of bright penumbral grains, chromospheric penumbral microjets \cite{2007Sci...318.1594K,tiwari15} and moving bright dots in the transition region \cite{2014ApJ...790L..29T} and in the corona \cite{alpert15}.
Observations with high cadence and high spatial resolution with ALMA have the potential to answer many open questions about these events, for instance about the origin of penumbral bright dots.  
In preparation of observations of active regions, a 3D modeling tool \cite[GX Simulator,][]{2014arXiv1409.0896N} has been developed, which can quickly compute synthetic maps in the ALMA spectral range based on photospheric input (white light and magnetogram) and chromospheric modelling constraints.

External triggering of sub-flares in braided coronal magnetic structures in active regions has been observed by Hi-C \citep{2013Natur.493..501C,2014SoPh..289.4393K,2014ApJ...795L..24T} and offers now an alternative to the spontaneous internal triggering of sub-flares \citep{1988ApJ...330..474P}, which is thought to be more common. 
ALMA can shed light on how important these different triggering mechanisms are and what role they play for coronal heating.

\paragraph{Coronal loops and coronal rain} 
\label{sec:rain}
Coronal loops, which are important building blocks in active regions, can  be subject to  thermal instabilities. 
The result is catastrophic cooling of the contained plasma from coronal temperatures down to chromospheric temperatures and with it the formation of dense blob-like structures (also known as ``coronal rain'') moving downwards along the loops \cite{2001ApJ...550.1036A,1970PASJ...22..405K,Leroy_1972SoPh...25..413L,Schrijver_2001SoPh..198..325S,2012ApJ...745..152A}. 
Coronal rain is observed commonly in active regions but appears in connection with coronal loops in general including post-flare loops. 
Observing coronal rain with ALMA can now help to further constrain the thermal characteristics of coronal/chromospheric loops. 
In particular, such observations may contribute to the currently ongoing debate concerning the substructure of loops and the question if the substructure is fully resolved with currently available (optical) instruments or not \cite{2012SoPh..280..457A,2012ApJ...745..152A,2014ApJ...797...36S,Antolin_etal_2015}.

\paragraph{Magnetic field measurements} 
ALMA observations in dual circular polarization mode will provide a powerful diagnostic tool for the magnetic field over a substantial range of chromospheric heights.  
The chromospheric magnetic field is fundamentally important but poorly known yet.  
The basic idea of measuring the magnetic fields of quiet Sun and non-flaring active regions from thermal bremsstrahlung was developed by \cite{1980SoPh...67...29B}  and  \cite{2000A&AS..144..169G} and tested on the RATAN-600 and NoRH data at short cm wavelengths. 
The proposed technique relies on the fact that the longitudinal component of the magnetic field can be derived from the observed degree of circular polarization and the brightness temperature spectrum.  
\cite{loukitchevabifrostmm2} carried out tests of this technique for ALMA frequencies using a snapshot from a 3D~radiation MHD simulation done with the Bifrost code \cite{2011AnA...531A.154G}, which represents a region of enhanced network. 
The magnetic field strength has a maximum of 2000\,G in the photosphere but rapidly decreases with height. 
The simulated circular polarization at 3\,mm falls within $\pm 0.5$\,\%, which results in the restored longitudinal magnetic field within $\pm 100$~G at the effective formation height. 
The method was found to reproduce the longitudinal magnetic field well with the ideal model data. 
However, further analysis, e.g. involving realistic noise, is needed to fully validate the approach. 
In a non-flaring active region, the degree of circular polarization reaches a few percent at mm wavelengths, which allows reliable determination of the longitudinal component of chromospheric magnetic field from ALMA observations. 
For more advanced active region magnetometry a combination of ALMA observational data with gyroresonant emission measurements in the microwave range available from the Very Large Array (VLA) and, in near future, from the Expanded Owens Valley Solar Array (EOVSA), the Siberian Solar Radio Telescope (SSRT), and the Chinese Solar Radio Telescope (CSRT) is desirable.

%================================================================================
\subsection{Prominences} 
\label{sec:promi}

ALMA offers the unique possibility to measure the kinetic temperature of the cool plasma ($T < 10,000$\,K) forming the dense cores of prominence fine structures with high spatial resolution. 
It will allow for investigating the elusive nature of prominence threads and for testing the hypothesis that prominence threads and coronal rain strands (see Sect.~\ref{sec:rain})  may be aspects of the same phenomenon occurring in magnetically different environments \citep{Lin_2011SSRv..158..237L,Tandberg-Hansen_1995ASSL..199.....T}. 
High-resolution ALMA observations of the prominence plasma could furthermore reveal the structure of the local magnetic field inside prominences, thus complementing present and future high-resolution spectro-polarimetric observations (e.g. those by DKIST (formerly ATST) or Solar-C). 
The plasma of quiescent prominence fine structures is mostly located in magnetic dips (see, e.g., \cite{2013A&A...551A...3G} or \cite{2013ApJ...766..126H}) and the information about the radius of curvature of the field inside the dips could provide an indirect test of the validity of an assumption of the force-free field in prominences.
The very high-resolution prominence visibility can be predicted by forward modelling based on 3D whole-prominence fine structure models like, e.g., developed by \cite{gunar2015}, which contains a realistic 3D prominence magnetic field configuration filled with numerous prominence plasma fine structures.
Prominences are discussed in detail by \cite{heinzel2015} who also considers predictions of the actual visibility of quiescent prominences and their fine structures in ALMA observations.

%================================================================================
%================================================================================
%================================================================================
\section{Conclusions and Outlook} 
%================================================================================

Solar observations with ALMA will produce without any doubt important results for a large range of topics which will advance our understanding of the atmosphere of our Sun. 
Many of these results will also have implications for other fields of astrophysics. 
For instance, a better understanding of chromospheric and coronal heating in the Sun is essential for understanding the heating of stellar atmospheres in general. 
It involves questions about the source of the so-called ``basal flux'' \cite{Schrijver1987,Perez2011} and the nature of a wide range of different activity levels among stars.  
Indeed, ALMA has already produced interesting results for other stars such as $\alpha$~Centauri \cite{2015A&A...573L...4L}.

The SSALMONetwork, which was introduced in this article, will --  in connection with the ongoing ALMA development studies and the relevant ARC nodes --  support and coordinate future activities in order to prepare solar observations with ALMA, to optimise them and to aid their interpretation. 
Another forthcoming publication \cite{ssalmon_ssrv15} will form the basis for expert teams, which will work on the scientific topics outlined in this article. 

\paragraph{Acknowledgements} 
S.~Wedemeyer acknowledges support (UiO-PES2020) by the Faculty of Mathematics and Natural Sciences of the University of Oslo, Norway, and the Research Council of Norway (grant 221767/F20).
M. B\'{a}rta thanks for the support of the European Commission through the CIG
grant PCIG-GA-2011-304265 (SERAF) and GACR grant 13-24782S. 
Roman Braj\v sa acknowledges support  from the European Commission FP7
projects eHEROES (284461, 2012-2015) and SOLARNET  (312495, 2013-2017), as
well as from the Croatian Science Foundation (project 6212 "Solar and
Stellar Variability").
M.~Barta, M.~Karlicky, E.~Kontar and V.~M.~Nakariakov acknowledge the Marie Curie PIRSES-GA-2011-295272 RadioSun project.
G.~Fleishman is supported by NSF grants AGS-1250374 and AGS-1262772 and NASA grant NNX14AC87G to the New Jersey Institute of Technology.
M. Loukitcheva acknowledges Saint-Petersburg State University for research grants 6.0.26.2010 and 6.37.343.2015, and grant RFBR 15-02-03835.

\bibliographystyle{elsarticle-num}

%================================================================================
%================================================================================
\end{document}